\documentclass[conference]{IEEEtran}
\usepackage{cite}

\ifCLASSINFOpdf
   \usepackage[pdftex]{graphicx}
  \DeclareGraphicsExtensions{.pdf,.jpeg,.png}
\else
\fi
\usepackage[inline]{enumitem}
\usepackage{amssymb}
\usepackage{booktabs}
\usepackage{amsmath}
\newcommand{\norm}[1]{\left\lVert#1\right\rVert}
\usepackage{url}

\begin{document}
\raggedbottom

%
\title{CAMBI: Contrast-aware Multiscale Banding Index}



%
\author{\IEEEauthorblockN{Pulkit Tandon\IEEEauthorrefmark{1}\textsuperscript{\textsection},
Mariana Afonso\IEEEauthorrefmark{2},
Joel Sole\IEEEauthorrefmark{2} and 
Luk\'{a}\v{s} Krasula\IEEEauthorrefmark{2}}
\IEEEauthorblockA{\IEEEauthorrefmark{1}Department of Electrical Engineering, 
Stanford University, CA, USA, 94305. tpulkit@stanford.edu}
\IEEEauthorblockA{\IEEEauthorrefmark{2}Netflix Inc., Los Gatos, CA, USA, 95032. \{mafonso, jsole, lkrasula\}@netflix.com\\
}
}

\maketitle
\begingroup\renewcommand\thefootnote{\textsection}
\footnotetext{Work done during an internship at Netflix.}
\endgroup


\begin{abstract}
Banding artifacts are artificially-introduced contours arising from the quantization of a smooth region in a video. Despite the advent of recent higher quality video systems with more efficient codecs, these artifacts remain conspicuous, especially on larger displays. In this work, a comprehensive subjective study is performed to understand the dependence of the banding visibility on encoding parameters and dithering. We subsequently develop a simple and intuitive no-reference banding index called CAMBI (Contrast-aware Multiscale Banding Index) which uses insights from Contrast Sensitivity Function in the Human Visual System to predict banding visibility. CAMBI correlates well with subjective perception of banding while using only a few visually-motivated hyperparameters.
\end{abstract}


%
\IEEEpeerreviewmaketitle

\section{Introduction}
\label{intro}

Banding artifacts are staircase-like contours introduced during the quantization of spatially smooth-varying signals, and exacerbated in the encoding of the video. These artifacts are visible in large, smooth regions with small gradients, and present in scenes containing sky, ocean, dark scenes, sunrise, animations, etc. Banding detection is essentially a problem of detecting artificially introduced contrast in a video. Even with high resolution and bit-depth content being viewed on high-definition screens, banding artifacts are prominent and tackling them becomes even more important for viewing experience. 

\begin{figure}
\centering
\includegraphics[width=3.4in]{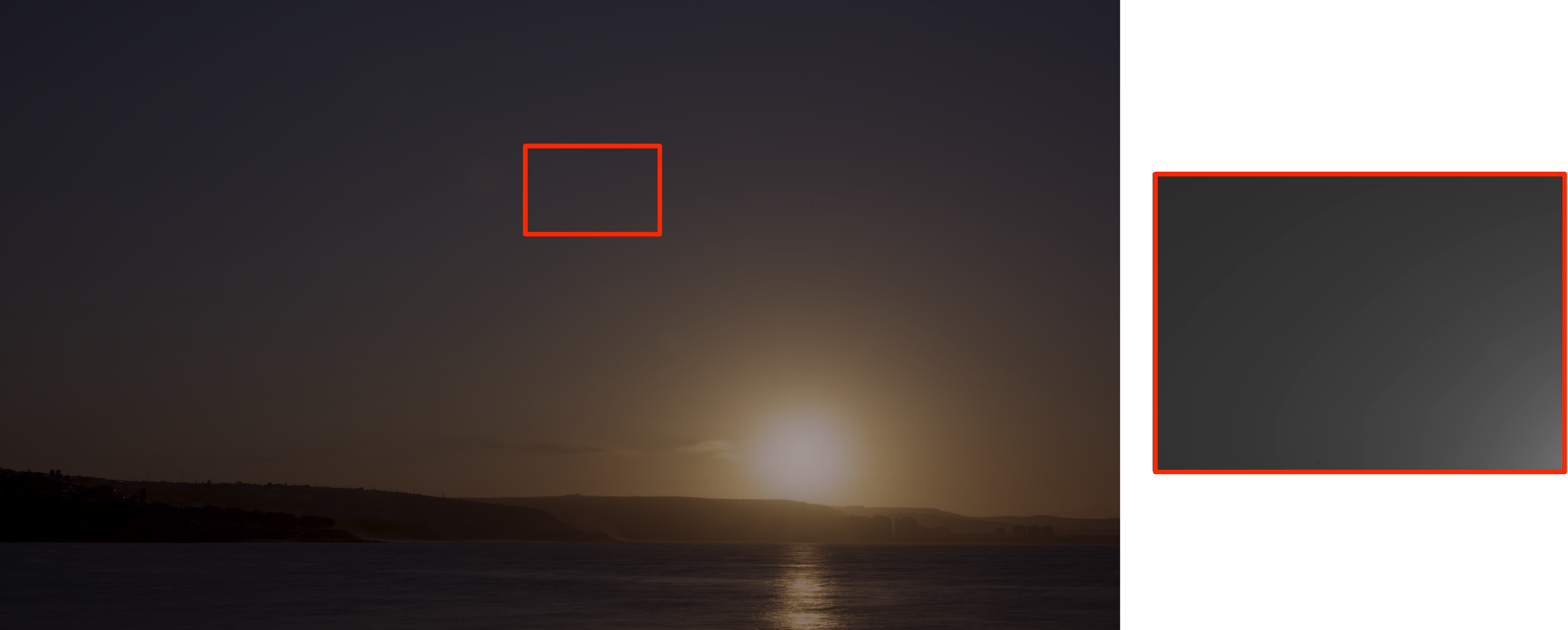}
\caption{Banding Motivation. Example from \textit{Night on Earth} series on Netflix (4k, 10-bit). Red box shows a zoomed-in luma segment with prominent bands.}
\label{fig_motivation}
\end{figure}

Figure \ref{fig_motivation} shows an example frame from \textit{Night on Earth} series on Netflix, encoded using a modern video codec, AV1 \cite{chen2018overview}, and the libaom encoder. Bands are clearly visible in the sky due to the intensity ramp present between the sun and its periphery. Traditional video quality metrics such as PSNR, SSIM \cite{wang2004image} or VMAF \cite{li2016toward} are not designed to identify banding and are hence not able to account for this type of artifact \cite{wang2016perceptual}, \cite{bband}, as we will also show in Section \ref{results}. These artifacts are most salient in a medium bitrate regime where the video is neither highly compressed and thus exacerbated by other artifacts, nor provided with large number of bits to faithfully represent the intensity ramp. Having a banding detection mechanism that is robust to multiple encoding parameters can help identify the onset of banding in the videos and serve as a first step towards its mitigation.

\textbf{Related Work.}
Although banding detection has been studied in the literature, no single metric or index is widely employed. Previous works on banding detection have focused on either false segment or false edge detection. For false segment detection, past methods have utilized segmentation approaches, such as pixel \cite{wang2016perceptual}, \cite{multiscale}, \cite{baugh2014advanced} or block-based segmentation \cite{jin2011composite}, \cite{wang2014multi}. For false edge detection, methods have utilized various local statistics such as gradients, contrast and entropy \cite{daly2004decontouring}, \cite{lee2006two}, \cite{huang2016understanding}. But both of these approaches suffer the hard problem of distinguishing between true and false segments/edges. Typically, this issue is solved by employing multiple hand-designed criteria obtained via observing a limited dataset. Moreover, most of these methods do not consider the possibility of dithering in the encoded video, which can be introduced by common tools such as ffmpeg \cite{tomar2006converting} during bit-depth reduction and can significantly affect the banding visibility. One recent no-reference banding detection method has outperformed previous work by using heuristics motivated by various properties of the Human Visual System, along with a number of pre-processing steps to improve banding edge detection \cite{bband}. This algorithm also contains a large number of hyperparameters trained and tested over a limited dataset \cite{wang2016perceptual}.

In this work we studied banding artifact's dependence on various properties of encoded videos, viz. quantization parameters, encoding resolution and incidence of dithering. We present a simple, intuitive, no-reference, distortion-specific index called \textbf{C}ontrast-\textbf{a}ware \textbf{M}ultiscale \textbf{B}anding \textbf{I}ndex (\textbf{CAMBI}), motivated by the algorithm presented in Ref. \cite{multiscale}. CAMBI directly tackles the problem of contrast detection by utilizing properties of Contrast Sensitivity Function (CSF) \cite{bovik2009essential}, instead of framing banding detection as a false segment/edge detection. In addition, CAMBI contains only few hyperparameters, most of which are visually-motivated. Results from the experiments conducted show that CAMBI has a strong linear correlation with subjective scores.

\section{Banding Detection Algorithm}

\begin{figure}
\centering
\includegraphics[width=3.4in]{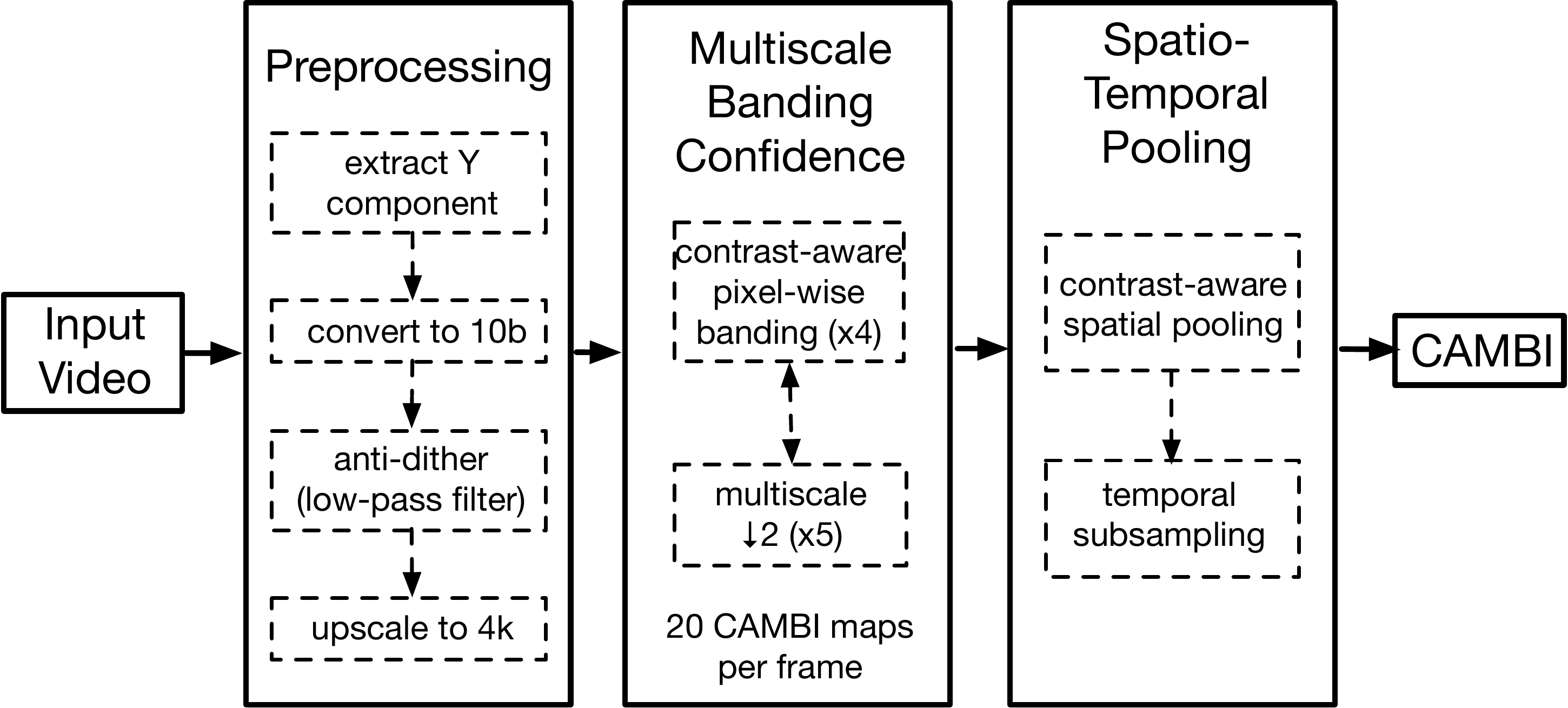}
\caption{Block diagram of the proposed algorithm.}
\label{fig_block}
\end{figure}

\begin{figure}
\centering
\includegraphics[width=3.4in]{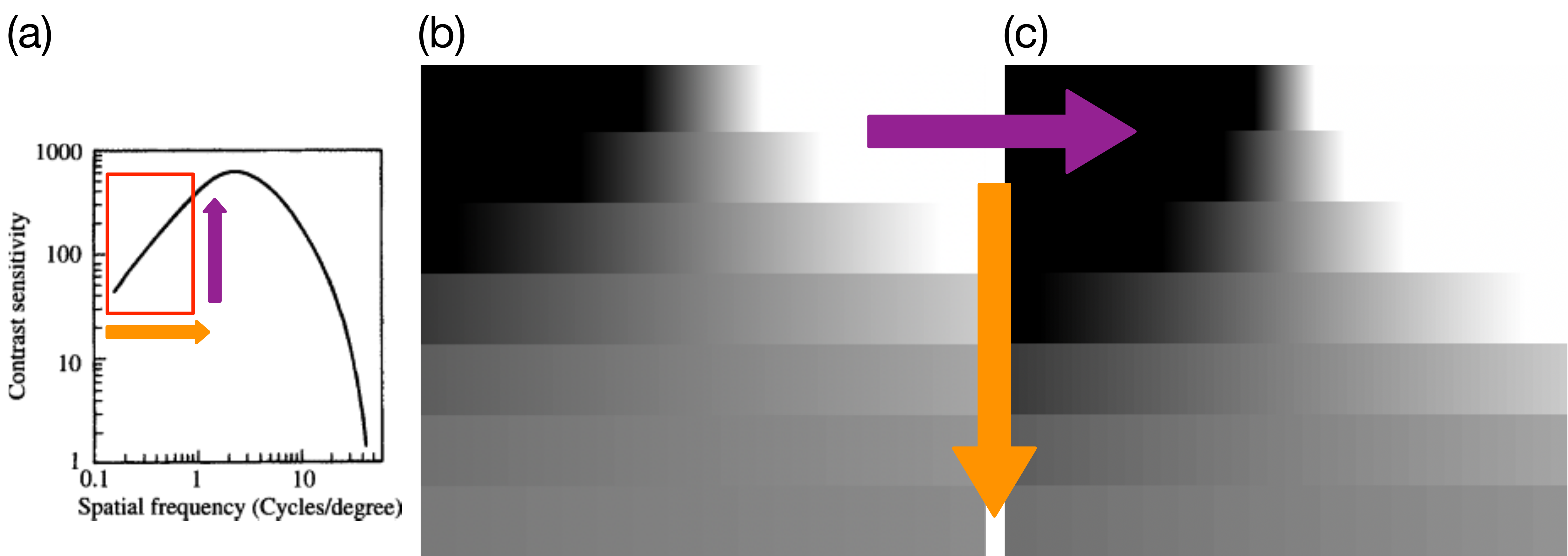}
\caption{Effect of Contrast Sensitivity Function (CSF) on banding visibility. \textit{(a)} CSF and its dependence on spatial frequency \cite{bovik2009essential}. \textit{(b) and (c)} Toy example showing banding visibility with smoothly varying intensity quantized at increasing contrast step (purple arrow) and spatial frequency (orange arrow).}
\label{fig_csf}
\end{figure}

We describe here the developed banding detection algorithm: \textbf{\textit{CAMBI}}. A block diagram describing all the steps involved in CAMBI is shown in Figure \ref{fig_block}. CAMBI operates as a no-reference banding detector. It takes a video as an input and produces a banding visibility score. CAMBI extracts multiple pixel-level maps at multiple scales, for temporally sub-sampled frames of the encoded video, and subsequently combines these maps into a single index motivated by human CSF \cite{bovik2009essential}. Steps involved in CAMBI are described next.

\subsection{Pre-processing}
Each input frame is taken through several pre-processing steps. Firstly, the luma component is extracted. Although it has been shown that chromatic banding exists, like most of the past works we assume that majority of the banding can be captured in the luma channel \cite{yoo2020gifnets}, \cite{denes2019visual}.

Next, dithering present in the frame is accounted for. Dithering is intentionally applied noise used to randomize quantization error, and has been shown to significantly affect banding visibility \cite{multiscale}, \cite{daly2004decontouring}. Presence of dithering makes banding detection harder, as otherwise clean contours might have noisy jumps in quantized steps, leading to unclean edges or segments detection. Thus to account for both dithered and non-dithered regions in a frame, we use a $2\times2$ averaging low-pass filter (\textit{LPF}) to smoothen the intensity values, in an attempt to replicate the low-pass filtering done by the human visual system. 

Low-pass filtering is done after converting the frame to a bit-depth of 10 (encodes studied in this work are 8-bit, but obtained from a 10-bit source as described in Section \ref{subjective_study}). This ensures that the obtained pixel values are in steps of one in 10-bit scale after application of LPF. Finally, we assume that the display is 4k (see Section \ref{subjective_study}), and hence irrespective of the encode resolution the frame is upscaled to 4k. Further steps in the algorithm are agnostic to the encode properties studied in this work, viz. resolution, quantization parameter, and incidence of dithering. Though we assume 10-bit sources and 4k displays in this work, CAMBI can be extended to encodes from sources at arbitrary bit-depths and display resolutions by modifying the bit-depth conversion and spatial upscaling steps appropriately. 

\subsection{Multiscale Banding Confidence}
\label{algo_main}

\begin{figure*}
\centering
\includegraphics[width=6.4in]{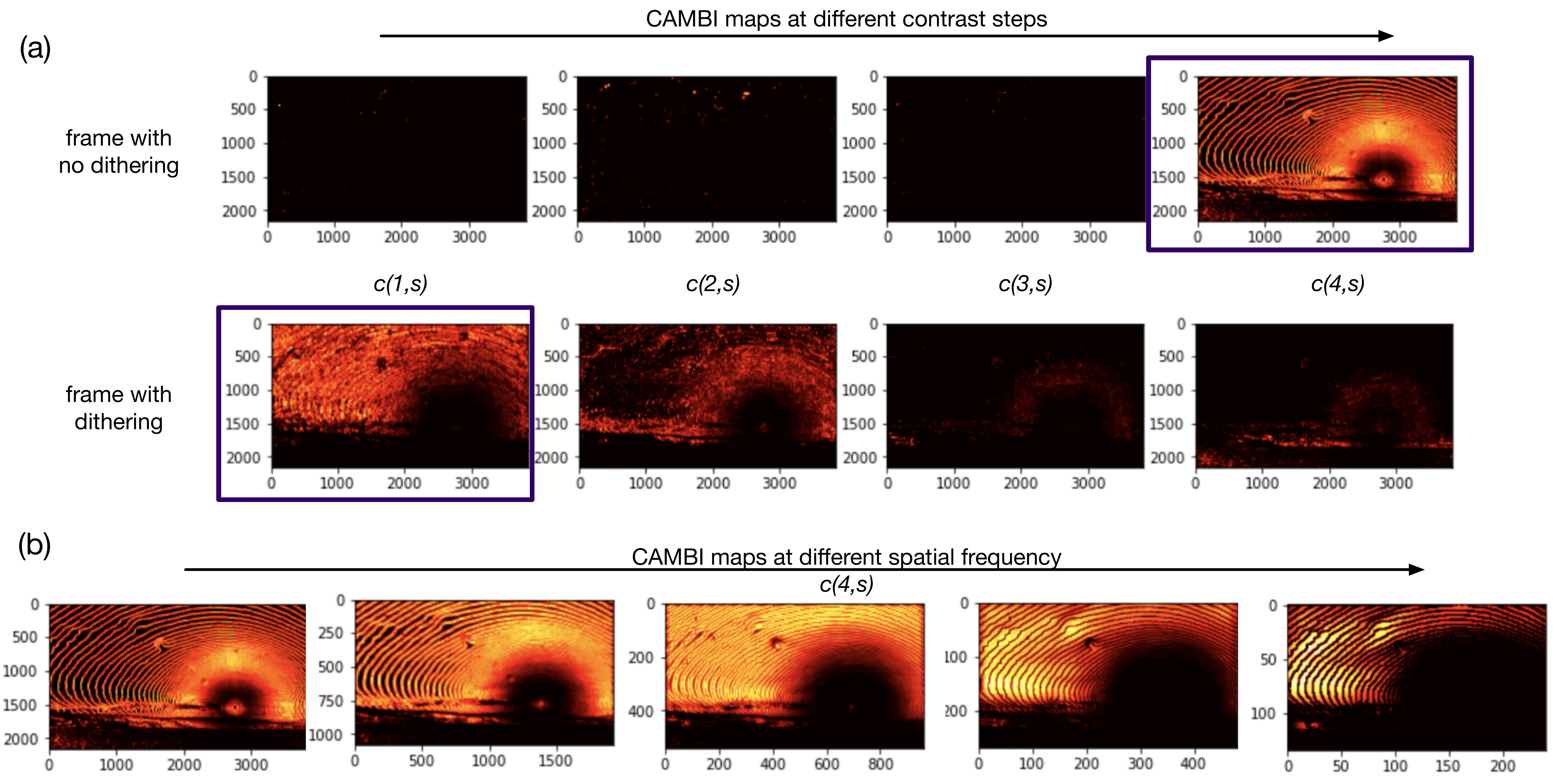}
\caption{Exemplary CAMBI maps. Frames are from example shown in Figure \ref{fig_motivation}. A warmer color represents higher banding confidence $c(k,s = 65\times65)$.}
\label{fig_CAMBI_maps}
\end{figure*}

As mentioned in Section \ref{intro}, we consider banding detection as a contrast-detection problem, and hence banding visibility is majorly governed by the CSF. CSF itself largely depends on the perceived contrast across the step and spatial-frequency of the steps, as illustrated in Figure \ref{fig_csf}. CAMBI explicitly tries to account for the contrast across pixels by looking at the differences in pixel intensity and does this at multiple scales to account for spatial frequency. 
    
CAMBI generalizes the approach used in \cite{multiscale}, which computes a pixel-wise banding confidence $c(s)$ at a scale $s$ as
\begin{equation}
\label{eq_c}
    c(s) = p(0,s)\times \max\left[\frac{p(-1,s)}{p(0,s)+p(-1,s)}, \frac{p(1,s)}{p(0,s)+p(1,s)}\right]
\end{equation}
where $p(k,s)$ is given by
\begin{equation}
\label{eq_pks}
    p(k,s)=
    \frac{\sum\limits_{\substack{\{(x',y')\in N_s(x,y)\mid\\\norm{\nabla(x',y')}< \tau_g\}}} \delta(I(x',y'),I(x,y)+k)}{\sum\limits_{\{(x',y')\in N_s(x,y)\mid\norm{\nabla(x',y')}< \tau_g\}} 1}
\end{equation}
In Eq. \ref{eq_pks}, $(x,y)$ refers to a particular pixel, and $I(x,y)$, $N_s(x,y)$ and $\norm{\nabla(x,y)}$ correspond to the intensity, neighborhood of a scale $s$ and gradient magnitude at this particular pixel, respectively. $\delta(.,.)$ is an indicator function. Thus, $p(k,s)$ corresponds to the fraction of pixel (in a neighborhood around  pixel $(x,y)$) with an intensity difference of $k$ amongst the set of pixels with gradient magnitude smaller than $\tau_g$. Hyperparameter $\tau_{g}$ ensures avoidance of textures during banding detection \cite{multiscale}. Therefore, Eq. \ref{eq_c} calculates a banding confidence $c(s)$ which explicitly tries to find if there is an intensity step of $\pm1$ in a pixel's non-texture neighborhood at scale $s$. $p(0,s)$ ensures that at the scale $s$, the pixel around which banding is being detected belongs to a visually-significant contour.
    
In CAMBI, the above approach is modified to explicitly account for multiple contrast steps and different spatial-frequencies, thus accounting for CSF-based banding visibility. This is done by calculating pixel-wise banding confidence $c(k,s)$ per frame at various different contrasts ($k$) and scales ($s$), each referred to as a \textit{\textbf{CAMBI map}} for the frame. A total of twenty CAMBI maps are obtained per-frame capturing banding across $4$ contrast-steps and $5$ spatial-frequencies.

For calculating CAMBI maps, Eq. \ref{eq_c} is modified as follows:
\begin{equation}
    \label{eq_c_CAMBI}
        c(k,s) = p(0,s)\max\left[\frac{p(-k,s)}{p(0,s)+p(-k,s)}, \frac{p(k,s)}{p(0,s)+p(k,s)}\right]
\end{equation}
where $k\in\{1,2,3,4\}$. Intensity differences of up to $\pm4$ are considered because of the conversion from 8-bit to 10-bit. If the pixel belongs to a dithering region it would have neighbouring pixels with intensity difference of $<$ $4$ because of the applied anti-dithering filter. On the other hand, if banding edge exists without any dithering in the frame, this would lead to an intensity difference of $\pm4$ at a bit-depth of 10, as a false contour appearing due to quantization will have pixels differing by $1$ on either side of the contour at bit-depth of 8. This leads to four CAMBI maps per frame, at each scale. Figure \ref{fig_CAMBI_maps}a shows the CAMBI maps obtained at different contrasts for the example shown in Figure \ref{fig_motivation}, for both dithered and non-dithered frame (see Section \ref{dataset}). Warmer colors represent higher $c(k,s)$ values and highlighted boxes clearly show that in undithered frame banding largely occurs at a contrast step of $k=4$ whereas for a frame containing dithering banding confidence shows up largely at lower contrast steps. 

To account for the banding visibility dependence on spatial frequency of bands, we modify the multiscale approach used by Ref. \cite{multiscale} to reduce the computational complexity. First, we fix the window-size ($s$) and then find $c(k,s)$ for frames after a mode-based downsampling is applied in powers of two from the initial resolution of 4k.  In total 5 scales are considered: 4k, 1080p, 540p, 270p and 135p. This leads to five CAMBI maps per frame at each contrast. Furthermore, a window-size ($s$) of $65\times65$ (centered at pixel) is chosen in this study which corresponds to $\sim1^{\circ}$ visual angle at 4k resolution based on subjective test design as described in Section \ref{subjective_study}. Thus, our multiscale approach calculates banding visibility at spatial-frequencies corresponding to visual degrees ($v^{\circ}$) of $\sim$\{$1^{\circ}$, $2^{\circ}$, $4^{\circ}$, $8^{\circ}$, $16^{\circ}$\}. Figure \ref{fig_CAMBI_maps}b shows CAMBI maps obtained at these five different scales at a contrast step of $4$, for a frame without dithering as shown in Figure \ref{fig_CAMBI_maps}a top panel. Figure \ref{fig_CAMBI_maps}b clearly shows that CAMBI is able to identify bands at various spatial-frequencies (e.g. high-frequency bands near sun at 4k and low-frequency bands away from the sun at 135p).

\subsection{Spatio-Temporal Pooling}
Finally, CAMBI maps obtained per frame are spatio-temporally pooled to obtain the final banding index. Spatial pooling of CAMBI maps is done based on the observation that above described CAMBI maps belong to the initial linear phase of the CSF (Figure \ref{fig_csf}, red box). Since perceived quality of video is dominated by regions with poorest perceived quality, only the worst $\kappa_{p}(p=30\%)$ of the pixels are considered during spatial pooling \cite{bband}. Though this improved correlation results (Section \ref{results}), using $\kappa_{p}(p=100\%)$ also leads to competitive correlation numbers (not shown). 
\begin{equation}
\label{eq_pool}
    \mbox{CAMBI$_f$} =\frac{ \sum\limits_{(x,y)\in\kappa_{p}}
    \sum\limits_{\substack{k=\\1,..,4}}
    \sum\limits_{\substack{v^{\circ}=\\1,2,..,16}}
    c(k,s)\times k \times \log_{2}\left(\frac{16}{v^{\circ}}\right)}
    {\sum\limits_{(x,y)\in\kappa_{p}}1}
\end{equation}
where $1/v^{\circ}$ represents spatial-frequency at which banding is detected (described in Section \ref{algo_main}).

Finally, CAMBI is applied to a frame every $\tau_s = 0.5$s and averaged, resulting in final CAMBI scores for the video. The value of $\tau_s$ was chosen based on temporal frequency dependence of CSF \cite{monaci2002color} as well as for implementation efficiency. According to our experiments, CAMBI was observed to be temporally stable within a single shot of a video but simple temporal pooling may fail if applied to a video with multiple shots. More sophisticated methods are planned for future work.
\begin{equation}
    \mbox{CAMBI} = \left. \sum\limits_{f \in \tau_s}\mbox{CAMBI$_f$} \middle/ \sum\limits_{f \in \tau_s}1 \right.
\end{equation}

\begin{table}
\renewcommand{\arraystretch}{1.3}
\centering
\caption{Hyperparameters used in CAMBI.}
\label{table_hyp}
\vspace{-0.1in}
\begin{tabular}{|c|c|}
\hline
low-pass filter (\textit{LPF}) & $2\times2$ avg filter\\
\hline
window-size ($s$ in $N_s$) & $65\times65$\\
\hline
gradient threshold ($\tau_g$) & $2$\\
\hline
spatial pooling ($\kappa_p$) & $30\%$\\
\hline
temporal sub-sampling ($\tau_s$) & $0.5$s\\
\hline
\end{tabular}
\vspace{-0.1in}
\end{table}

Hyperparameters used in CAMBI are summarized in Table \ref{table_hyp} and validation results are shown in Section \ref{results}.

    
    

\section{Performance Evaluation}

\subsection{Banding Dataset}
\label{dataset}
A banding dataset was created for this study based on existing Netflix catalogue. Nine 4k-10bit source clips with duration between 1 and 5 seconds were utilized. Of these, eight clips had various levels of banding and one had no banding. Nine different encodes were created for each of these sources by using the following steps: 
\begin{enumerate*}
    \item downsampling source to appropriate resolution (1080p, quad-HD or 4k) and bit-depth (8-bit) using ffmpeg, 
    \item encoding the downsampled content at three different QPs (12, 20, 32) using libaom.
\end{enumerate*}
Ordered-dithering gets introduced in the frames during downsampling by ffmpeg and gets selectively pruned during encoding (dependent on QP and resolution). Thus, we also added a tenth encode per source where dithering is not introduced to explicitly validate whether CAMBI can track banding visibility across dithering. This encode was done at maximum quality (4k resolution, 12 QP) to juxtapose the banding visibility in absence of dithering against other encoding parameters.

\subsection{Subjective Study}
\label{subjective_study}
The subjective evaluation was performed on the above described dataset by asking viewers familiar with banding artifacts to rate the worst-case annoyance caused by the banding across all video frames on a modified DCR scale from 0 (unwatchable) to 100 (imperceptible) \cite{recommendationitu}. For each viewer, six encodes of an additional (not from banding dataset) source content with expected score ranging from $0$ to $100$ (in steps of $20$) were firstly shown, along-with the expected scores, in a training session. Following this, a single-stimulus test with randomized exhaustive clip-order from the banding dataset was performed remotely\footnote{Future reader, note that we are in the middle of a pandemic.}. Each viewer watched the test content on a 15 inch Apple Macbook Pro. All videos were played in a loop until terminated by the viewer. In addition, viewers were asked to maintain a distance of $\sim1.5\times$screen height from the screen throughout the experiment. All the encodes presented were 8-bit, upsampled to 4k and cropped to $2880\times1800$ pixels for uniformity across subject's display resolutions. Although no attempt was made to control ambient lightning, we asked the viewers to adjust the display brightness to around 80\% of the maximum. A detailed study including ambient lightning dependence is planned for future work.

\begin{figure}
\centering
\includegraphics[width=3.4in]{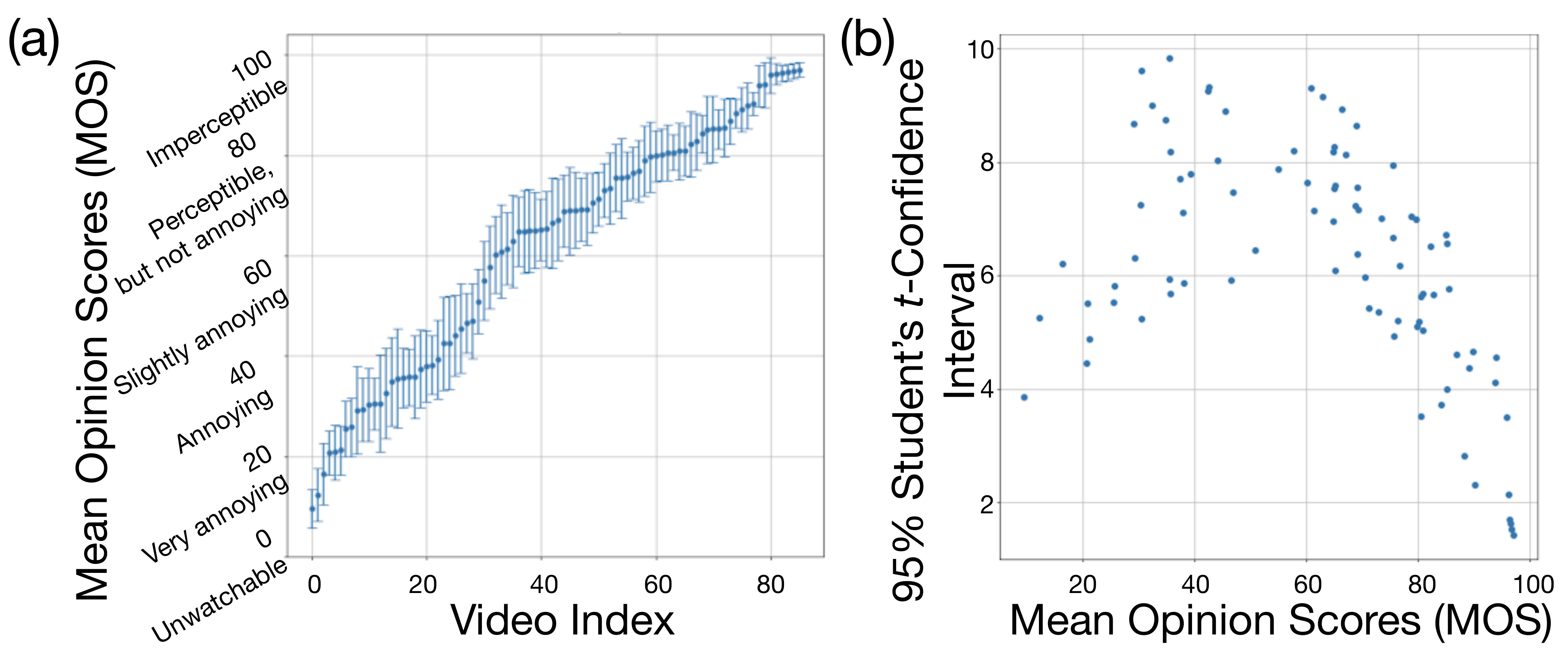}
\caption{Subjective test properties (86 encodes, 23 viewers). \textit{(a)} Designed test had a thorough coverage of the scale, and \textit{(b)} mean opinion scores obtained had a $95\%$ Student's \textit{t}-confidence interval of $<10$.}
\label{fig_st}
\end{figure}

A total of $86$ encodes were evaluated in this study (with four 4k sequences removed because of non-real time decoding of highest quality AV1 encode by browser). All chosen sequences had qualities of VMAF $>80$ and PSNR $>40$ dBs, highlighting the problem of banding prevalence even in highly-rated videos using traditional metrics. To the best of our knowledge, this subjective study is the first to account for dependence of banding on resolution and presence of dithering. In total $23$ subjects participated in this study. Figure \ref{fig_st} shows that the banding scores obtained had a thorough coverage of the DCR scale as well as a $95\%$ Student's \textit{t}-confidence interval of $<$$10$.

\begin{figure}
\centering
\includegraphics[width=3.4in]{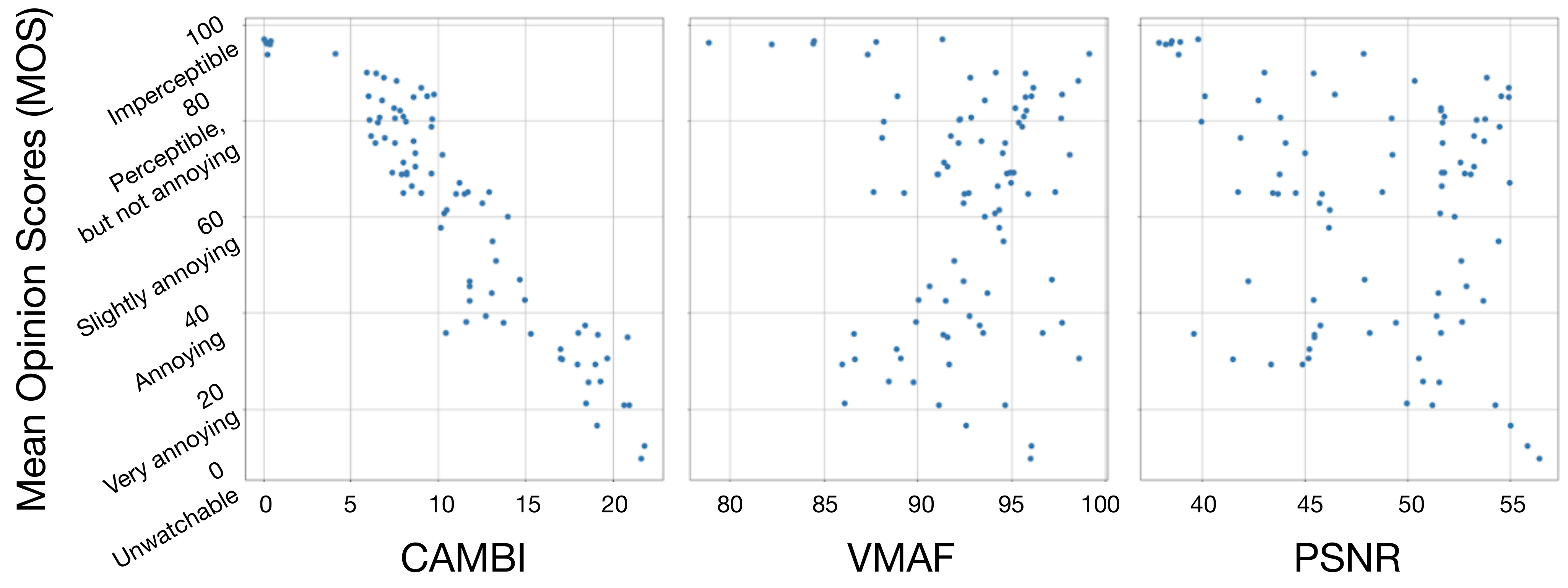}
\caption{Subjective study results. \textit{(left panel)} CAMBI is linearly correlated with mean opinion scores (MOS) obtained through subjective study, \textit{(middle, right panels)}  whereas VMAF and PSNR are uncorrelated with the MOS.}
\label{fig_results}
\end{figure}

\subsection{Results}
\label{results}
\subsubsection{\textbf{CAMBI is linearly correlated with subjective scores}}
The Mean Opinion Scores (MOS) obtained from the subjective study were compared with the output from CAMBI and two objective quality metrics, VMAF and PSNR. Results are shown in Figure \ref{fig_results}. We can see that CAMBI provides high negative correlation with MOS while VMAF and PSNR have very little correlation. A number of correlation coefficients, namely Spearman Rank Order Correlation (SROCC), Kendall-tau Rank Order Correlation (KROCC), Pearson's Linear Correlation (PLCC) and $(|$KROCC$|+1)/2$ over statistically-significant pairs (C0) \cite{lukas_c0} are reported in Table \ref{table_results}. From a total of $3655$ comparisons possible amongst MOS of $86$ videos, $2895$ pairs had a difference in MOS which was statistically significant and $95\%$ of these orderings were correctly identified by CAMBI. Individual scores reported are mean $\pm$ standard deviation (maximum) correlation coefficients when an individual viewer's subjective scores are compared against MOS and suggests CAMBI performs equivalent to an individual sensitive in identifying banding. These results suggest that CAMBI is able to accurately estimate banding visibility across a number of variables with high linear-dependence (without any additional fitting).

\begin{table}
\renewcommand{\arraystretch}{1.3}
\centering
\caption{Performance comparison of metrics against subjective scores.}
\vspace{-0.1in}
\label{table_results}
\begin{tabular}{ccccc}
    \toprule
    & CAMBI $\downarrow$ & VMAF $\uparrow$ & PSNR $\uparrow$ & Individual\\
    \midrule
    SROCC & -0.923 & 0.088 & -0.202 & 0.844 $\pm$ 0.108 (0.953)\\
    KROCC & -0.765 & 0.099 & -0.124 & 0.678 $\pm$ 0.108 (0.821)\\
    PLCC & -0.929 & 0.000 & -0.271 & 0.853 $\pm$ 0.097 (0.957)\\
    C0 & 0.950 & 0.545 & 0.409 & ---\\
    \bottomrule
\end{tabular}
\vspace{-0.1in}
\end{table}

\subsubsection{\textbf{CAMBI is unbiased over range of video qualities}}
\begin{figure}
\centering
\includegraphics[width=2.4in]{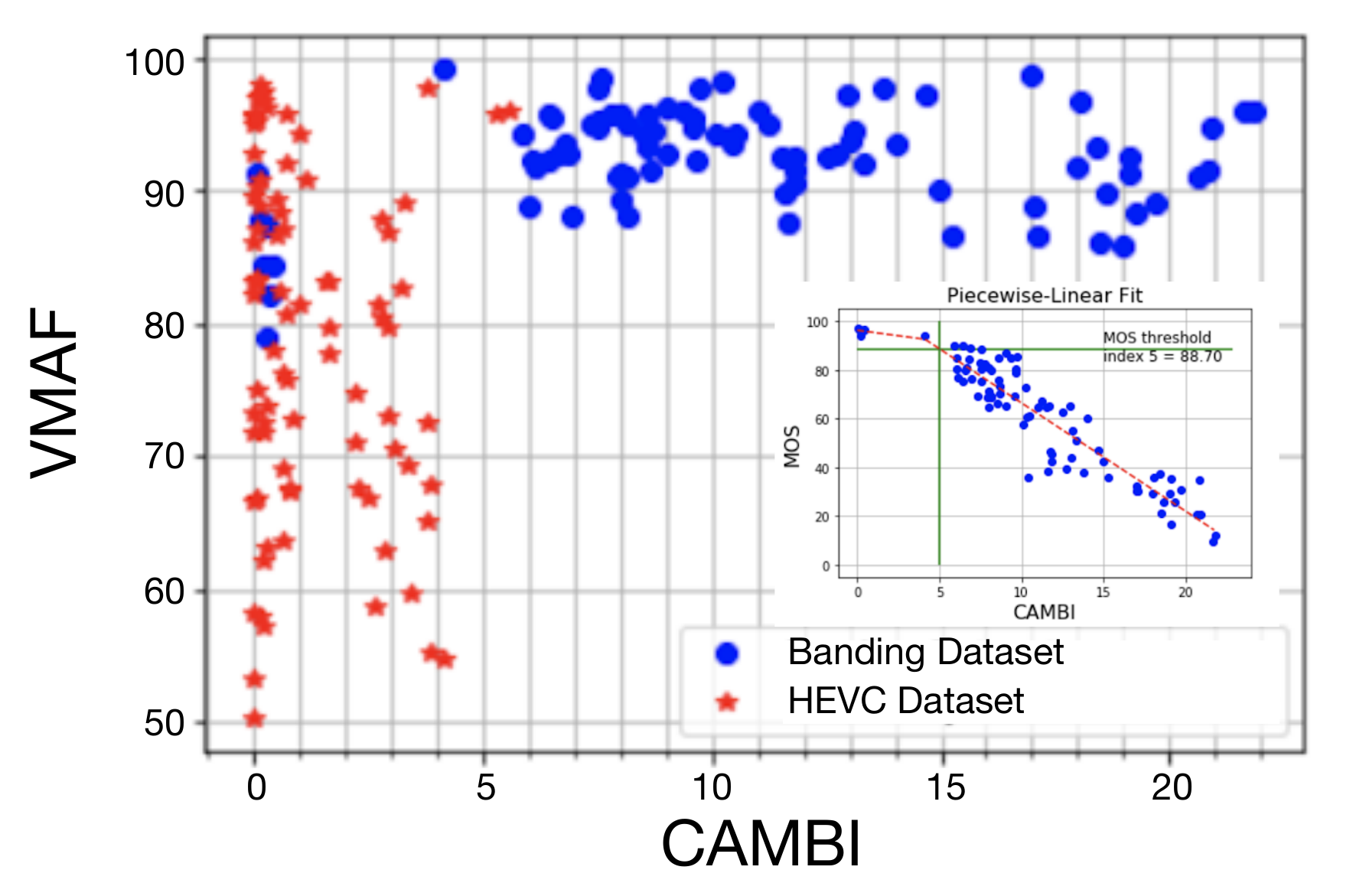}
\caption{Checking for False Positives. CAMBI when applied to another dataset with no banding \cite{HEVC_dataset} doesn't over-predict banding scores. Inset shows a piecewise linear fit between MOS and CAMBI.}
\label{fig_false_positives}
\end{figure}

CAMBI was also validated on an independent dataset without visible banding artifacts. This dataset contains 84 HEVC encodes from seven 4k 10-bit sources with a range of VMAF scores \cite{HEVC_dataset}. Figure \ref{fig_false_positives} shows CAMBI against VMAF for both the datasets. Though CAMBI is designed for worst-case banding visibility and verified using subjective scores based on worst-case annoyance caused by banding, this false-positive analysis seems to indicate that CAMBI does not over-predict banding scores. Figure \ref{fig_false_positives} also provides an interpretation for the range of CAMBI scores, where CAMBI $<5$ would suggest no visible banding artifacts are present.

\section{Conclusion}
In this work, we present a simple and intuitive, no-reference, distortion-specific banding index called \textbf{\textit{CAMBI}}. CAMBI is able to estimate banding visibility across multiple encoding parameters by employing visually-motivated computational motifs. We conducted a comprehensive subjective study to validate CAMBI and showed that it has a high correlation and a near-linear relationship with mean opinion scores. In addition, the small number of hyperparameters and false-positive analysis suggest a good generalizability of this index. In the future, we plan to validate and improve CAMBI on a larger subjective dataset using videos with varied bit-depths and encoded using different video codecs. CAMBI can also be used in conjunction with or integrated as an additional feature in future versions of VMAF, and to aid the development of debanding algorithms.


\section*{Acknowledgment}
The authors would like to thank Zhi Li, Christos Bampis and codec team at Netflix for feedback on this work and all the observers who participated in the subjective test.



\bibliographystyle{IEEEtran}
\bibliography{citations.bib}

\begin{thebibliography}{10}
\providecommand{\url}[1]{#1}
\csname url@samestyle\endcsname
\providecommand{\newblock}{\relax}
\providecommand{\bibinfo}[2]{#2}
\providecommand{\BIBentrySTDinterwordspacing}{\spaceskip=0pt\relax}
\providecommand{\BIBentryALTinterwordstretchfactor}{4}
\providecommand{\BIBentryALTinterwordspacing}{\spaceskip=\fontdimen2\font plus
\BIBentryALTinterwordstretchfactor\fontdimen3\font minus
  \fontdimen4\font\relax}
\providecommand{\BIBforeignlanguage}[2]{{%
\expandafter\ifx\csname l@#1\endcsname\relax
\typeout{** WARNING: IEEEtran.bst: No hyphenation pattern has been}%
\typeout{** loaded for the language `#1'. Using the pattern for}%
\typeout{** the default language instead.}%
\else
\language=\csname l@#1\endcsname
\fi
#2}}
\providecommand{\BIBdecl}{\relax}
\BIBdecl

\bibitem{chen2018overview}
Y.~Chen, D.~Murherjee, J.~Han, A.~Grange, Y.~Xu, Z.~Liu, S.~Parker, C.~Chen,
  H.~Su, U.~Joshi \emph{et~al.}, ``An overview of core coding tools in the av1
  video codec,'' in \emph{2018 Picture Coding Symposium}.\hskip 1em plus 0.5em
  minus 0.4em\relax IEEE, 2018.

\bibitem{wang2004image}
Z.~Wang, A.~C. Bovik, H.~R. Sheikh, and E.~P. Simoncelli, ``Image quality
  assessment: from error visibility to structural similarity,'' \emph{IEEE
  transactions on image processing}, vol.~13, no.~4, pp. 600--612, 2004.

\bibitem{li2016toward}
Z.~Li, A.~Aaron, I.~Katsavounidis, A.~Moorthy, and M.~Manohara, ``Toward a
  practical perceptual video quality metric,'' \emph{The Netflix Tech Blog},
  vol.~6, p.~2, 2016.

\bibitem{wang2016perceptual}
Y.~Wang, S.-U. Kum, C.~Chen, and A.~Kokaram, ``A perceptual visibility metric
  for banding artifacts,'' in \emph{2016 IEEE International Conference on Image
  Processing (ICIP)}.\hskip 1em plus 0.5em minus 0.4em\relax IEEE, 2016, pp.
  2067--2071.

\bibitem{bband}
Z.~Tu, J.~Lin, Y.~Wang, B.~Adsumilli, and A.~C. Bovik, ``Bband index: a
  no-reference banding artifact predictor,'' in \emph{ICASSP 2020-2020 IEEE
  International Conference on Acoustics, Speech and Signal Processing
  (ICASSP)}.\hskip 1em plus 0.5em minus 0.4em\relax IEEE, 2020, pp. 2712--2716.

\bibitem{multiscale}
S.~Bhagavathy, J.~Llach, and J.~Zhai, ``Multiscale probabilistic dithering for
  suppressing contour artifacts in digital images,'' \emph{IEEE Transactions on
  Image Processing}, vol.~18, no.~9, pp. 1936--1945, 2009.

\bibitem{baugh2014advanced}
G.~Baugh, A.~Kokaram, and F.~Piti{\'e}, ``Advanced video debanding,'' in
  \emph{Proceedings of the 11th European Conference on Visual Media
  Production}, 2014, pp. 1--10.

\bibitem{jin2011composite}
X.~Jin, S.~Goto, and K.~N. Ngan, ``Composite model-based dc dithering for
  suppressing contour artifacts in decompressed video,'' \emph{IEEE
  Transactions on Image Processing}, vol.~20, no.~8, pp. 2110--2121, 2011.

\bibitem{wang2014multi}
Y.~Wang, C.~Abhayaratne, R.~Weerakkody, and M.~Mrak, ``Multi-scale dithering
  for contouring artefacts removal in compressed uhd video sequences,'' in
  \emph{2014 IEEE Global Conference on Signal and Information Processing
  (GlobalSIP)}.\hskip 1em plus 0.5em minus 0.4em\relax IEEE, 2014, pp.
  1014--1018.

\bibitem{daly2004decontouring}
S.~J. Daly and X.~Feng, ``Decontouring: Prevention and removal of false contour
  artifacts,'' in \emph{Human Vision and Electronic Imaging IX}, vol.
  5292.\hskip 1em plus 0.5em minus 0.4em\relax International Society for Optics
  and Photonics, 2004.

\bibitem{lee2006two}
J.~W. Lee, B.~R. Lim, R.-H. Park, J.-S. Kim, and W.~Ahn, ``Two-stage false
  contour detection using directional contrast and its application to adaptive
  false contour reduction,'' \emph{IEEE Transactions on Consumer Electronics},
  vol.~52, no.~1, pp. 179--188, 2006.

\bibitem{huang2016understanding}
Q.~Huang, H.~Y. Kim, W.-J. Tsai, S.~Y. Jeong, J.~S. Choi, and C.-C.~J. Kuo,
  ``Understanding and removal of false contour in hevc compressed images,''
  \emph{IEEE Transactions on Circuits and Systems for Video Technology},
  vol.~28, no.~2, pp. 378--391, 2016.

\bibitem{tomar2006converting}
S.~Tomar, ``Converting video formats with ffmpeg,'' \emph{Linux Journal}, vol.
  2006, no. 146, p.~10, 2006.

\bibitem{bovik2009essential}
\BIBentryALTinterwordspacing
K.~Seshadrinathan, T.~N. Pappas, R.~J. Safranek, J.~Chen, Z.~Wang, H.~R.
  Sheikh, and A.~C. Bovik, ``Chapter 21 - image quality assessment,'' in
  \emph{The Essential Guide to Image Processing}, A.~Bovik, Ed.\hskip 1em plus
  0.5em minus 0.4em\relax Boston: Academic Press, 2009, pp. 553 -- 595.
  [Online]. Available:
  \url{http://www.sciencedirect.com/science/article/pii/B9780123744579000214}
\BIBentrySTDinterwordspacing

\bibitem{yoo2020gifnets}
I.~Yoo, X.~Luo, Y.~Wang, F.~Yang, and P.~Milanfar, ``Gifnets: Differentiable
  gif encoding framework,'' in \emph{Proceedings of the IEEE/CVF Conference on
  Computer Vision and Pattern Recognition}, 2020.

\bibitem{denes2019visual}
G.~Denes, G.~Ash, H.~Fang, and R.~K. Mantiuk, ``A visual model for predicting
  chromatic banding artifacts,'' \emph{Electronic Imaging}, vol. 2019, no.~12,
  pp. 212--1, 2019.

\bibitem{monaci2002color}
G.~Monaci, G.~Menegaz, S.~Susstrunk, and K.~Knoblauch, ``Color contrast
  detection in spatial chromatic noise,'' blah, Tech. Rep., 2002.

\bibitem{recommendationitu}
Recommendation, ``Itu-tp. 913,'' \emph{ITU}, 2016.

\bibitem{lukas_c0}
L.~Krasula, K.~Fliegel, P.~Le~Callet, and M.~Kl{\'\i}ma, ``On the accuracy of
  objective image and video quality models: New methodology for performance
  evaluation,'' in \emph{2016 Eighth International Conference on Quality of
  Multimedia Experience (QoMEX)}.\hskip 1em plus 0.5em minus 0.4em\relax IEEE,
  2016, pp. 1--6.

\bibitem{HEVC_dataset}
\BIBentryALTinterwordspacing
Netflix, \emph{Test Conditions}, accessed 2020. [Online]. Available:
  \url{http://download.opencontent.netflix.com/?prefix=Netflix_test_conditions/}
\BIBentrySTDinterwordspacing

\end{thebibliography}
%



  

\end{document}